\documentclass[prl,twocolumn,showpacs,preprintnumbers,aps,floatfix,showpacs]{revtex4}
\usepackage{epsfig,bm,epsf,graphics}
\usepackage{dcolumn}

\newcommand{\be}{\begin{equation}}
\newcommand{\ee}{\end{equation}}
\newcommand{\ben}{\begin{eqnarray}}
\newcommand{\een}{\end{eqnarray}}

\begin{document}


\title{Interacting black holes on the brane: the seeding of binaries}
\author{A. S. Majumdar}
\email{archan@bose.res.in}
\author{Anita Mehta}
\email{anita@bose.res.in}
\affiliation{S. N. Bose National Centre for Basic Sciences, Block JD, Salt
Lake, Kolkata 700 098, India}
\author{J. M. Luck}
\email{luck@spht.saclay.cea.fr}
\affiliation{Service de Physique Th\'eorique (URA 2306 of CNRS),
CEA Saclay, 91191 Gif-sur-Yvette cedex, France}

\date{\today}

\begin{abstract}
We consider the evolution of sub-horizon-sized black holes which are formed
during the high energy phase of the braneworld scenario. These black holes
are long-lived due to modified evaporation and accretion of
radiation during the radiation dominated era. We argue
that an initial mass difference between any two neighbouring 
black holes is always
amplified because of their exchange of energy with the surrounding radiation. 
We present a scheme of binary formation based on mass differences suggesting
that such a scenario could lead to binaries with observable signatures.

\end{abstract}

\pacs{98.80.Cq}


\maketitle

Cosmology in the braneworld scenario has inspired much activity in recent
times. The feasibility of a large extra spatial dimension is the cornerstone
of the Randall-Sundrum (RSII) braneworld model\cite{randall} in which
all the standard model fields are confined to our observable $3$-brane, except
for gravity which can propagate also in the bulk. The braneworld description
of our universe entails an early high energy
phase during which the evolution of the universe is significantly altered.
A particular feature of interest in RSII cosmology is the
evolution of primordial $5$-dimensional black
holes. Black holes forming out of
the collapse of horizon-sized density perturbations during the high
energy phase obey a different evaporation law\cite{guedens} compared to
$4-$dimensional black holes. It has been shown recently that
accretion from the surrounding radiation bath can dominate over evaporation
during the high energy phase\cite{majumdar}. This leads to prolonged
survival of these primordial black holes with multifarious cosmological and
astrophysical consequences. There exists the possibility that
these black holes could be a significant fraction of cold dark matter.
Several observational constraints on primordial black
holes get modified in the braneworld scenario, as has been shown
recently\cite{clancy}.

Black holes as dark compact objects in galaxy haloes are the
target of many recent and ongoing ob\-ser\-va\-tions\cite{popowski}.
A large number
of these black holes are expected to be in the form of binaries. Several
interesting observational evidences
ranging from lensing effects to hypervelocity star
ejections have
been put forward to bolster claims in support of black hole
binaries\cite{yu} in a wide mass spectrum. Recently, it has been proposed
that sub-lunar mass binaries comprised of braneworld black holes could
produce detectable gravitational waves in their coalescing stage\cite{inoue}.
Gravitational lensing experiments do indeed leave open the possibility
of the existence of sub-lunar mass black holes in certain mass
ranges\cite{alcock}. A formation mechanism for binaries
in standard cosmology based on the inhomogeneous spatial distribution of
primordial black holes has been proposed\cite{nakamura}.

In this Letter we present a scheme of binary formation for primordial
braneworld black holes based on mass differences. We consider
sub-horizon sized black holes in the RSII model formed during the high energy
phase whose evolution has been studied in Refs.\cite{guedens,majumdar}.
We shall consider the interaction of two neighbouring
black holes which are initially well separated for their gravitational
attraction to be approximated by Newtonian dynamics. The physical distance
between
two such neighbours increases with the Hubble expansion of the high energy
radiation dominated era. These
black holes exchange energy via the processes of evaporation and accretion
with the radiation bath, and through it, with each other.
The initial mass ratio of two such black
holes is on average proportional to the square of the ratio of their
formation times\cite{guedens}. We will see from the evolution that the
initial mass difference between two such black
holes always increases during the radiation dominated era. We will
then argue that such mass differences will facilitate the formation of binaries
during the standard low energy phase via three-body gravitational
interactions. We will consider an example of sub-lunar mass binary\cite{inoue}
formation through this scheme.

Let us first briefly state some essential results of
Refs.\cite{guedens,majumdar}. The mass of a primordial braneworld black hole
formed at time $t_0$ with initial mass $m_0$ during the high energy phase
($t < t_c$ with $t_c \equiv l/2$) grows due to accretion (evaporation is
negligible, except for black holes with very small initial mass, or very
low accretion efficiency $f$, i.e.,
except for $(M_0/M_4)^2 < 2A/(2B-1)(t_0/t_c)$, where $A \simeq 3/((16)^3\pi)$,
and $B = 2f/\pi$, with $0 \le f \le 1$) such that
\be
\label{1}
{m(t) \over m_0} \simeq \Biggl({t\over t_0}\Biggr)^B
\ee
This equation holds up to $t\simeq t_c$ assuming radiation
domination persists, i.e., the radiation density $\rho_R=(3m_4^2)/[32\pi (t_c +
t) t]$
stays greater than the sum of the energy density in all black
holes, $\rho_{BH}$.
The formation time is related to the initial mass by
\be
\label{2}
{t_0\over t_4} \simeq {1\over 4}\Biggl({m_0\over
m_4}\Biggr)^{1/2}\Biggl({l\over l_4}\Biggr)^{1/2}
\ee
with the $4$-dimensional Planck mass (time) labelled as $m_4$ ($l_4$).
Assuming accretion to be effective only during the high energy phase,
these $5$-dimensional black holes evaporate as
\be
\label{3}
m(t) = \biggl(m_\mathrm{max}^2 - 2Am_4^2{t\over t_c}\biggr)^{1/2}
\ee
during the standard low energy regime
$t \gg t_c$, with $m_\mathrm{max}$ being the maximum mass a black hole
reaches via accretion. [Recently, there has been a conjecture of rapid
evaporation through
conformal modes for large braneworld black holes with size greater then
the AdS radius $l$\cite{tanaka}. However, our focus will be on much smaller
black holes.]

Now let us consider the evolution of two neighbouring black holes,
BH1 and BH2, having
masses $m_1$ and $m_2$ respectively, with initial masses $m_{1_0} < m_{2_0}$,
or $t_{1_0} < t_{2_0}$.
BH2 forms at a physical distance $d_0$ from BH1. Their physical separation
$d$ grows at a rate $\propto t^{1/4}$. We shall assume that
$d_0 > l$, and that the gravitational potential
$\phi = [m/(m_4^2d)][1+(2l^2)/(3d^2)] \ll 1$ so that exchange of energy
between them via gravitational waves can be neglected
during the high energy regime.
The change of mass of each black hole is effected by the
net sum of its Hawking evaporation depending on its temperature, and
the accretion of radiation depending on the mean radiation density in the
universe\cite{majumdar}. Moreover,
due to the presence of BH2, the net rate of change of the mass $\dot{m}_1$ of
BH1 gets a further contribution proportional to the product of $\dot{m}_2$ and
the solid angle subtended by BH2 on BH1, and vice-versa. In other words,
BH1 feels the local difference from the mean radiation density as a result of
an accreting (or evaporating) BH2, and similarly for BH2. Thus the evolution
equation for each of them can be written as
\be
\dot{m}_i = {Bm_i\over t} - {Am_4^2\over m_it_c} - {r_i^2\dot{m}_j\over 4d^2}
\label{evolution}
\ee
for $i,j=1,2$ [$(r_i/l_4) = (8/3\pi)^{1/2}(l/l^4)^{1/2}(m_i/m_4)^{1/2}$
is the Schwarzschild radius for $5$-dimensional black holes\cite{guedens}].
The first and second terms on the r.h.s represent accretion from the average
radiation density,
and evaporation respectively\cite{majumdar}. The last term arises from
the local
inhomogeneity in radiation density due to the $j$-th black hole. Considering
both BH1 and BH2 to be in their accreting phases at the formation time
$t_{0_2}$
of BH2, it is
possible to see from Eqs.(\ref{evolution}) that for BH1 the first term is
comparable to the interaction (third) term when
$(d/r_2) \sim ([t-t_{1_0}]/[t-t_{2_0}])^{1/2}$. Thus, for a while after
its formation, BH2 is able to suppress the growth of BH1. If on the other
hand, BH2 forms at a time when BH1 begins to evaporate, with or without the
aid of the interaction, BH2 registers enhanced growth due to the locally
denser radiation coming from the evaporating BH1.

The effect of two interacting black holes on their evolution can be more
precisely formulated as follows. Defining $\tilde{t}=Am_4^2t/t_c$,
Eqs.(\ref{evolution}) can be written as
\be
\dot{m}_i = {Bm_i\over \tilde{t}} -{1\over m_i} - g{m_i\dot{m}_j\over
\tilde{t}^{1/2}}
\label{evolve2}
\ee
where
\be
g = {4A^{1/2}\over 3\pi}\biggl({l_4\over d_0}\biggr)^2\biggl({t_0\over
t_4}\biggr)^{1/2}\biggl({t_c\over t_4}\biggr)^{1/2}
\label{coupling}
\ee
can be regarded as a coupling constant between BH1 and BH2 dependent
on their initial separation $d_0$. Now, in terms of the variables
$s=\ln(\tilde{t}/t_0)$ and $x_i=m_i/\tilde{t}^{1/2}$,
Eq.(\ref{evolve2}) can be cast in an autonomous form, i.e.,
\be
x_i' = \biggl[B - \frac{1}{2} -g\biggl(\frac{x_j}{2} - x'_j\biggr)\biggr]x_i -
\frac{1}{x_i}
\label{autonom}
\ee
where accents denote derivatives with respect to $s$.
In the following we confine our attention to small enough values of $g$,
such that $g^2x_1x_2<1$.
This is indeed the condition for Eq.(\ref{autonom}) to be well-behaved.
For two equal mass black holes ($x_1=x_2=x$),
this condition becomes $g<g_\star=1/x$.
This limit coupling strength in fact corresponds to the limit of coalescence,
since the
black hole radius $r$ is of the order of the separation $d$ for $g=g_\star$.
Below the critical value $g_c = [2(2B-1)^3/27]^{1/2}$
of the coupling constant,
the dynamics of two equal mass black holes has a stable fixed point $x_F$.
Our interest
is in two unequal mass black holes which are initially well separated for
gravitational interaction to be subdominant.
If one considers a small initial mass difference,
then the two black hole masses will evolve as
$x_1=x(s) - \epsilon(s)$ and $x_2=x(s)+\epsilon(s)$,
with $\epsilon(s) \approx \epsilon(0)\exp\left(\int_0^s\Lambda(x(u))du\right)$,
where the instantaneous Lyapunov exponent
\be
\Lambda(x) = \frac{2B-1}{2} + \frac{1+2gx +Bg^2x^4}{x^2(1-g^2x^2)}
\label{lyapunov}
\ee
is larger than the constant $(2B-1)/2$.
It follows that
any small initial mass difference blows up exponentially in the
first stages of dynamical evolution.
In particular, the symmetric fixed point $x_F$ is always linearly
unstable against a perturbation of the form $\delta x_1 = -\delta x_2$.

It is thus clear that the physical solutions of Eqs.(\ref{evolve2}) exhibit
disequilibration for black holes masses, i.e., if for two neighbouring
black holes, $m_1(t_0) < m_2(t_0)$, then $m_1(t) \ll m_2(t)$ for $t>t_0$.
Requiring the
era $t$ with a number density $n_{BH}$ of black
holes of average mass $m(t)$ to be radiation dominated, i.e.,
$\rho_{BH} \equiv mn_{BH} \le \rho_R$, the mean strength of the coupling
$\bar{g}$ between two such
black holes should be such that
\be
\bar{g} \le {A^{1/2}\over 4}\biggl({m_4\over
m_0}\biggr)^{2/3}\biggl({t_4^2\over t_0t_c}\biggr)^{1/6} \ll g_\star
\label{avcoupl}
\ee
In general however, there could be spatial inhomogeneities in the formation
of black holes. The growth rate of any given black hole is hence impacted
by the mass spectrum and spatial distribution of all neighbouring black holes.
For a system of two black holes BH1 and BH2 formed at eras $t_{1_0}$ and
$t_{2_0}$ with the initial mass difference $(\Delta_{12})_0 = m_{2_0} -
m_{1_0}$, and when
both are in their accreting phases, the mass difference at time $t \gg t_{2_0}$
is given by (from Eqs.(1) and (2))
\ben
(\Delta_{12})(t) - (\Delta_{12})_0 \ge \nonumber \\
m_{2_0}\Biggl[\Biggl({t\over t_{2_0}}\Biggr)^B\Biggl(1-
\biggl({m_{1_0}\over m_{2_0}}\biggr)^{1-B/2}\Biggr) - \Biggl(1-{m_{1_0}\over
m_{2_0}}\Biggr)\Biggr]
\label{massdiff}
\een
with the equality sign holding for $g=0$, i.e., when the interaction term
can be neglected.

For the present analysis, we are assuming that black hole accretion is
restricted to the high energy phase. Hence, when the universe exits
the high energy phase, the masses of BH1 and BH2 are frozen to $m_1$
and $m_2$ (say). Hawking evaporation continues to be negligible for a long
time beyond $t_c$\cite{majumdar}. Consider a spherical region of radius $d$ in
which the average matter density (contributed by BH1 and BH2) is given by
\be
\bar{\rho}_{BH} = {3(m_1+m_2)\over 4\pi d^3}
\label{avmatt}
\ee
Since radiation density in this region falls off as $a^{-4}$, the sphere
of radius $d$ will become matter dominated at a time $t_f$ given from Eqs.(1)
and (2) by
\be
t_f \simeq 2\biggl({8\over 3\pi}\biggr)^3\biggl({d_0\over
r_{2_0}}\biggr)^6\biggl({t_0\over t_c}\biggr)^{2B+1/2}t_c
\label{formtime}
\ee
For $t>t_f$ the matter dominated region with radius $d$ gets cut off
from the background radiation dominated expansion which continues up to the
era of matter-radiation equality $t_{eq} \gg t_c$. For $t>t_f$, the black
holes BH1 and BH2 will form a bound system\cite{nakamura}.
Assuming that a bound system (or binary) gets formed only in the low
energy regime ($t_f>t_c$), it follows from Eq.(\ref{formtime}) that
\be
{d_0\over r_{2_0}} > {1\over 64}\biggl({3\pi\over
8}\biggr)^{1/2}\biggl({t_c\over t_{2_0}}\biggr)^{{4B+1\over 12}}
\label{distrad}
\ee
With the initial density fraction $\alpha_{m_0} (= \rho_{BH}(t_0)/\rho_R(t_0)$)
for black holes formed with
mass $m_0$, the mean initial separation is
$\bar{d_0}=(3\pi/8)^{1/2}(r_0/\alpha^{1/3}_{m_0})$.
[Two black holes with initial separation less than that given in
Eq.(\ref{distrad}) will
form a bound system in the high energy phase.
Accretion of radiation will be negligible after binary formation, with a
resultant smaller lifetime. In the present analysis we shall not consider
details of such a scenario.]
The size of
this bound system $d_f$ is related to the initial black hole separation
by $d_f = d_0(t_c/t_0)^{1/4}(t_f/t_c)^{1/2}$. Using Eqs.(1), (2) and
(\ref{formtime}),
\be
d_f = {1\over 16}\biggl({t_c\over t_0}\biggr)^{3-B}\biggl({d_0\over
l}\biggr)^3d_0
\label{formdist}
\ee
with its mean value given by $\bar{d}_f =
(4/\alpha^{4/3}_{m_0})(t_0/t_c)^{1+B}l$.

Binary formation proceeds as a result of 3-body interaction. Let us choose
three neighbouring black holes to have mass diffrences at $t_c$ given by
$m_2 - m_1=\Delta_{12}$ and $m_3 - m_2=\Delta_{23}$.   Such a scenario
will arise generically since the initial mass differences are due to horizon 
sized
collapse at different eras, and possible interactions through the radiation
background can only amplify such mass differences between neighbouring
black holes during the high energy regime.
BH2 will become gravitationally bound to BH3 at time $t_f$ with an elliptical
orbit whose major axis
\be
a_f \propto d_f
\label{majax}
\ee
The role of BH1 is to provide the tidal
force required to prevent head-on collision between BH2 and BH3. The minor
axis $b_f$ of the binary is
proportional to the tidal force times the squared free fall
time\cite{nakamura}, i.e.,
\be
b_f \propto 2\biggl({m_2 - \Delta_{12}\over 2m_2 +\Delta_{23}}\biggr)
\biggl({d_f\over d_P}\biggr)^3a_f
\label{minax}
\ee
where $d_P$ is the perpendicular distance of BH1 from the mid-point of
the BH2-BH3 axis.
Although BH2 binds to BH3 at time $t_f$, mass differences hamper the
possibility of BH1 being gravitationally bound to BH3. Even
if the three black hole separations are equal before binary formation, the
total region encompassing the binary and BH1 will become matter dominated at a
time $(t_{md})_{13}$ much later than $t_f$, given by
$(t_{md})_{13} \ge t_f[(m_{3_0}+m_{2_0})/(m_{1_0}+m_{2_0})]^{11/4-B}$,
(since using Eqs.(2) and (11), $t_f \propto m_0^{B-11/4}$)
with the equality sign signifying $g=0$. But during the interval
$t_f<t<(t_{md})_{13}$, BH1
continues to move away from the BH2-BH3 binary with the Hubble flow, i.e.,
$d_P$ registers growth by a factor $[(t_{md})_{13}/t_f]^{1/2}$.
The gravitational interaction between the binary and BH1 at $t = (t_{md})_{13}$
is much weakened compared to the strength of
the BH2-BH3 interaction.

The eccentricity of the binary orbit can be obtained from Eqs. (\ref{majax})
and (\ref{minax}). For uniformly distributed black holes formed around
the era $t_0$ with masses $\sim m_0$, the mean eccentricity is given by
\be
\bar{e} = \Biggl[1 - \eta^2 2\biggl({m_2-\Delta_{12}\over
2m_2+\Delta_{23}}\biggr)^2\Biggr]^{1/2}
\label{eccent}
\ee
where $\eta$ is an $O(1)$ geometrical factor\cite{nakamura}.
The Newtonian approximation that we have used for describing the gravitational
interaction between black holes forming a binary is valid if the gravitational
potential $|\phi| \ll 1$, which for the BH2-BH3 binary reads $(m_1/m_4^2) \ll
d_f(1-e)$. Using Eqs.(1), (\ref{formdist}) and (\ref{eccent}) one can verify
that the Newtonian analysis holds good for $(t_4/t_0)^{2B-1} \ll
\eta^2\alpha_{m_0}^{-4/3}[(m_2-\Delta_{12})/2m_2+\Delta_{23})]^2$.

As a particular example of the scenario outlined by us, consider three black
holes formed about the era $t_0 \sim 10^{26}t_4$ with average mass $m_0 \sim
10^{24}m_4$, radius $r_0 \sim 10^{27}l_4$ and $\Delta \sim O(m_1)$.
We have chosen $l= 10^{30}l_4$, about
the maximum allowed by present experimental bounds. Setting the accretion
factor $B=1/2$, one gets $m_\mathrm{max} \simeq m(t_c) = 10^{26}m_4 \sim
10^{21}\mathrm{g}$. The closure bound for a population of
black holes with $m_0 \sim 10^{24}m_4$ can be obtained using the prescription
of Clancy et al.\cite{clancy}, and turns out to be $\alpha_{m_0} < 6\times
10^{-18}$ for the initial energy fraction. Choosing the maximum allowed value
of $\alpha_{m_0}$ the mean initial
separation  is $\bar{d}_0 \sim 10^6r_0 \sim 1\mathrm{cm}$.
Nevertheless, the validity of the
Newtonian approximation $|\phi| \ll 1$ is readily ensured for a choice
$d_0 \sim 10^3r_0 $ for the specific system of three black
holes we are considering for binary formation. For these parameters,
binary formation
takes place at $t_f \sim 10^{12}t_c$, with the major axis
$a_f \sim 10^7\mathrm{cm}$.
Note here that we have estimated the mean values for the above
parameters assuming a uniform distribution of black holes. 
Inhomogeneities in the spatial distribution for black holes would
change these values on a case-to-case basis. However, at least some
of the binaries would have parameters close to the mean values.
These would then fall in the category of sub-lunar compact
objects discussed in Ref.\cite{inoue}, and gravitational waves emitted
during their coalescence are within the observational range of third
generation gravitational wave detectors.

To summarize, we have studied the effect of interaction of horizon-sized
black holes formed during the radiation dominated high energy phase of the
RSII braneworld scenario. We have seen that the processes of black hole
evaporation and accretion mediated via the radiation bath causes the mass
difference between two neighbouring black holes to be always amplified
during the high energy era. We have presented a scheme by which binary
formation takes place through three-body gravitational interaction during
the standard low energy era. This scheme  of binary formation based on mass
differences could be viewed as an additional possible mechanism to
the scheme based on spatial
inhomogeneities\cite{nakamura}. In a realistic scenario, both spatial and
mass inhomogeneities are expected in any given configuration of black
holes, and these separate effects will interfere to either abet or inhibit
binary formation depending upon the particular configuration. The specific
example we have furnished suggests that the parameters for such binaries 
might lie within the range of
sub-lunar compact objects amenable for detection by forthcoming
observations of gravitational waves such as by the EURO detector\cite{inoue}.
It is important
to note that we have
considered black hole accretion only during the high energy phase. If
accretion of radiation continues during the standard
low energy regime up to $t_\mathrm{eq}$, as claimed in
Refs.\cite{majumdar2}, it might
be worthwhile to investigate the possible formation of more massive binaries
through primordial black holes. One should keep in mind that black holes also
grow by accreting mass from the surrounding matter distribution in the matter
dominated era. Finally, the possibility of accreting the homogeneous
and presently dominating cosmological or scalar field energy\cite{bean},
if realized, could push up the masses further.


\begin{thebibliography}{99}
\bibitem{randall}
L. Randall and R. Sundrum, Phys. Rev. Lett. {\bf 83}, 4690 (1999).
\bibitem{guedens}
R. Guedens, D. Clancy and A. R. Liddle, Phys. Rev. D{\bf 66}, 043513 (2002).
\bibitem{majumdar}
A. S. Majumdar, Phys. Rev. Lett. {\bf 90}, 031303 (2003);
R. Guedens, D. Clancy and A. R. Liddle, Phys. Rev. D{\bf 66}, 083509 (2002).
\bibitem{clancy}
D. Clancy, R. Guedens and A. R. Liddle, Phys. Rev. D{\bf 68}, 023507 (2003);
Y. Sendouda, S. Nagataki and K. Sato, Phys. Rev D{\bf 68}, 103510 (2003).
\bibitem{popowski}
See for example, EROS collaboration, Astron. Astrophys. {\bf 400}, 951 (2003);
P. Popowski et al., astro-ph/0304464; J. Ziolkowski, astro-ph/0307307;
P. A. Charles and M. J. Coe, astro-ph/0308020.
\bibitem{yu}
Q. Yu, MNRAS {\bf 331}, 935 (2002); M. Milosavljevic and D. Marritt, Astrophys.
J. {\bf 596} (2003); Q. Yu and S. Tremaine, Astrophys. J. {\bf 599}, 1129
(2003).
\bibitem{inoue}
K. T. Inoue and T. Tanaka, Phys. Rev. Lett. {\bf 91}, 021101 (2003).
\bibitem{alcock}
C. Alcock et al., Astrophys. J. {\bf 499}, L9 (1998); G. F. Marani et al.,
Astrophys. J. {\bf 512}, L13 (1999).
\bibitem{nakamura}
T. Nakamura, M. Sasaki, T. Tanaka and K. S. Thorne, Astrophys. J.{\bf 487},
L139 (1997); K. Ioka, T. Chiba, T. Tanaka and T. Nakamura, Phys. Rev.
D{\bf 58}, 063003 (1998).
\bibitem{tanaka}
T. Tanaka, Prog. Theor. Phys. Suppl. {\bf 148}, 307 (2003); R. Emparan, A.
Fabbri and N. Kaloper, JHEP {\bf 0208}, 043 (2002); R. Emparan, J.
Garcia-Bellido and N. Kaloper, JHEP {\bf 0301}, 079 (2003).
\bibitem{majumdar2}
A. S. Majumdar, P. Das Gupta and R. P. Saxena, Int. J. Mod. Phys. D{\bf 4},
517 (1995); N. Upadhyay, P. Das Gupta and R. P. Saxena, Phys. Rev. D{\bf 60},
063513 (1999); P. S. Custodio and J. E. Horvath, Phys. Rev. D{\bf 58}, 023504
(1998); Phys. Rev. D{\bf 60}, 083002 (1999).
\bibitem{bean}
R. Bean and J. Magueijo, Phys. Rev. D{\bf 66}, 063505 (2002).

\end{thebibliography}
\end{document}